# THE EXISTENCE OF BARYONS AT $z = 1000$*

DOUGLAS SCOTT & MARTIN WHITE

*Center for Particle Astrophysics, University of California,*
*Berkeley, CA 94720-7304*

## ABSTRACT

Fluctuations in the microwave background have now been detected over a wide range of angular scales, and a consistent picture seems to be emerging. The data cannot currently be used to constrain a large number of cosmological parameters, but it is clear that there is more information than just the normalization of the models. Here we use the data to constrain a second parameter, namely the amplitude of the Doppler peak, which we do using a phenomenological approach to the radiation power spectrum. We find that the data prefer a Doppler peak of height $\simeq 3$ (above a flat line normalized to be unit height), with a purely flat spectrum ruled out at the 95% confidence level. Although there are concerns about foregrounds and the possibility of non-Gaussian fluctuations, we believe that the existence of a peak at degree scales is established by the data. This immediately implies that reionization was unimportant for the microwave background. It also potentially leads to difficulties for models where the fluctuations were produced by topological defects. Independent constraints on $\Omega_B$, on the slope $n$, etc. will need to wait for further data. At the moment, the simple presence of a Doppler peak should be seen as strong supporting evidence for standard dark matter-dominated models with some few percent of baryons at $z \simeq 1000$.

*Subject headings:* cosmic background radiation — cosmology: theories and observations — Doppler peaks: proved

## 1. Introduction

The existence of microwave background fluctuations over a wide range of angular scales has now been firmly established (see ref. 1 for a review), and emphasis is shifting toward studies that try to extract cosmological information from the experimental data. There have been several papers that combine the data from two experiments, usually the COBE DMR results on the largest scales plus a specific smaller angular-scale experiment, to place constraints on some cosmological parameters or models[2,3,4,5,6,7,8]. Some authors have even considered the results from many experiments, but with no definitive conclusions[9,10,11].

We believe that there are now enough independent experimental measurements of microwave background anisotropies at different scales that it is possible to combine the available data to obtain a robust answer to a relatively modest question.

---

* *or* "How High are the Doppler Peaks, and Other Tall Tales of the CMB"

Instead of trying to rule out specific cosmological models we take a more phenomenological approach. Firstly, we set up a 'toy-model' for the radiation power spectrum which is flat on large angular-scales and has a peak in power around multipole $\ell \simeq 250$. This increase at sub-degree scales corresponds to the so-called Doppler peaks in standard dark matter power spectra. Secondly, we take the data from the different experiments and convert them into a measure of power through each window function, so that they can all be plotted together for comparison, and so that they can be combined to place constraints. Finally we calculate the best-fitting height for the Doppler peak in our phenomenological power spectrum.

We find that a totally flat scale-invariant spectrum is ruled out by the data (at the 95% confidence level), which instead prefer some sort of Doppler peak with height $\simeq 3$ relative to the Sachs-Wolfe part of the radiation power spectrum. This result is remarkably close to what theorists had been anticipating, and it has some immediate implications for cosmology.

## 2. The Radiation Power Spectrum

It has become standard practice in CMB anisotropy studies to work in terms of the multipole moments of the temperature anisotropy. One conventionally defines $C_\ell = \left\langle |a_{\ell m}|^2 \right\rangle$, where $\Delta T/T(\theta,\phi) = \sum_{\ell m} a_{\ell m} Y_{\ell m}(\theta,\phi)$, with $Y_{\ell m}$ the spherical harmonics, and where the angled brackets represent an average over the ensemble of possible fluctuations. Assuming that the fluctuations are gaussian distributed, the models are uniquely specified by giving their $C_\ell$'s, which are usually plotted as $\ell(\ell+1)C_\ell$ vs. $\ell$. This is the power per logarithmic interval in $\ell$, or a 2D power spectrum on the sphere. Such a plot usually starts as a flat line at small $\ell$, corresponding to Sachs-Wolfe fluctuations in an $n = 1$ spectrum of matter fluctuations, with $P_{\text{matter}}(k) \propto k^n$.

The model $C_\ell$'s are obtained numerically from integrating the coupled Boltzmann equations for each fluid. It has become apparent[10,12,13] that variations caused by different cosmological parameters are not 'orthogonal', in the sense that essentially identical $C_\ell$'s can be found for different sets of parameters. Attempts to extract such parameters from CMB data are further complicated by the fact that theories only predict the expectation values of the $a_{\ell m}$'s (or the $C_\ell$'s) for an ensemble of skies[14,15,16,17], so that there is an unavoidable level of uncertainty when comparing with observations. These problems may be somewhat overcome however by using non-CMB constraints (see for example the plot of the matter power spectrum in ref. 1).

In order to formulate a question that the data may already be able to answer, we resort to some theoretical prejudice and fall back on some assumptions that have been common in such studies. We will assume that the power spectrum of radiation fluctuations is at least phenomenologically similar to that obtained from models like the 'standard' Cold Dark Matter model, although we need not assume that all the dark matter is cold. Specifically we assume that the power spectrum is flat (corresponding to $n = 1$) on the largest scales, that $\Omega_0 = 1$, and that the tensor-mode (gravity wave) contribution is small (i.e. $T/S \ll 1$). Under these assumptions the most prominent feature of theoretical power spectra is the rise above the flat line from about $\ell \simeq 100$.

Doppler peaks are the name generally used to describe the bumps and wiggles in the radiation power spectrum at $\ell$'s of a few hundred.* They are caused by the oscillations of the baryon-photon fluid before the Universe recombined. The different peaks and troughs correspond to photon density and velocity perturbations which have had integral number of half oscillations before entering the Jeans scale, with complications caused by the dark matter potential wells and the thickness of the last scattering surface. Higher $\Omega_B$ will generally lead to a smaller Jeans length, allowing perturbations to grow more before coming inside this scale and starting to oscillate. The oscillations will therefore be of greater amplitude for higher $\Omega_B$, leading to higher Doppler peaks when the photons are last scattered. The exact heights of the various bumps and wiggles comes from a combination of adiabatic and velocity effects and so depend on the specifics of the cosmological model (for example the height of the first peak is fairly insensitive to $h$, while the relative heights of subsidiary peaks have quite a strong $h$ dependence). However, experiments are sensitive to a wide range of $\ell$, which will somewhat wash out these variations. Moreover, the position of the first Doppler peak depends essentially only on the geometry of the Universe. The scale is determined by the Jeans length at last scattering, which subtends an angle corresponding to $\ell \sim 250\,\Omega_0^{-1/2}$ for standard recombination at $z \simeq 1100$. So for an $\Omega_0 = 1$ model and no significant reionization (our assumptions), the position of the main Doppler peak is well-determined. The damping scale of the $C_\ell$'s is also a fairly robust physical quantity. It is determined by the Silk damping scale and the thickness of the last scattering surface. Although there is some dependence on cosmological parameters the damping scale will be roughly $\ell \simeq 1500$.

In order to keep our power spectrum simple we approximate the Doppler peaks by a single peak. Specifically we take a phenomenological power spectrum of the form

$$\ell(\ell+1)C_\ell = 6C_2 \left\{1 + \frac{A_D}{1+y(\ell)^2}\right\} \Big/ \left\{1 + \frac{A_D}{1+y(2)^2}\right\}, \qquad (1)$$

---

* The naming of the Doppler peaks seems to be due to Bond & Efstathiou, who used it in talks etc. from the mid-80's, although they appear never to have quite referred to the name in print until much later![10,?] The actual origin of the peaks can be traced back (in less and less familiar-sounding language) through seminal papers such as Wilson & Silk[18], Doroshkevich, Zel'dovich & Sunyaev[19] and Peebles & Yu[20]. A general understanding of the importance of oscillations in the photon-baryon fluid goes back even earlier, e.g. to Silk[21] or Sakharov[22]. Indeed some Russians authors[23] refer to the Doppler peaks as Sakharov oscillations. However, it is clear that the peaks could not be fully understood with the large- and small-angle approximations used by early workers in this field; it wasn't until the detailed numerical calculations of Bond & Efstathiou[24], complete with the language of $P_{\rm rad}(k)$ and the $C_\ell$'s, that what we understand by the Doppler peaks had really been described. What you call them is rather a matter of taste. The first peak actually comes from the photon monopole term at last scattering, not the velocity term, so perhaps adiabatic peak would be more accurate (see ref. 25). However, the term 'Doppler' peaks seems to have become common in the literature, whether it is an accurate name or not, c.f. 'planetary' nebulae!

with
$$y(\ell) = \frac{\log_{10} \ell - 2.4}{0.38},$$
i.e. a constant plus a Lorentzian, with the amplitude at $\ell = 2$ divided out so that $A_D$ is the height above the Sachs-Wolfe plateau. The parameters for the center and width of the Lorentzian were fitted to accurate $C_\ell$'s for a standard CDM model ($h = 1/2, \Omega_B = 0.06$) provided by N. Sugiyama (see e.g. refs. 26,27). This fit is shown in Fig. 1, where the CDM model is the solid line, the best Lorentzian curve is the short-dashed line, and the best Gaussian curve is the long-dashed line. In fact the curves were fitted to points equally spaced in $\log \ell$, chosen from the CDM model. The easiest way to do this was just to take points at $\ell = 2^N$ with $N = 1\ldots 10$, which are the filled circles in Fig. 1. You can see that a Gaussian is not a very satisfactory fit, but that the Lorentzian curve is surprisingly accurate.

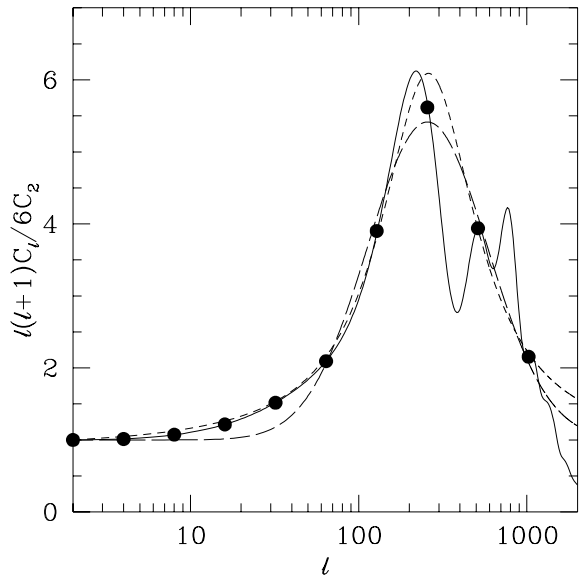

**Figure 1:** Our phenomenological fit to the $C_\ell$'s is illustrated here for a standard CDM model with $\Omega_B = 0.06$ and $h = 0.5$ (solid line). The points are simply the values $\ell = 2^N$ for $N = 1\ldots 10$ (an easy way of choosing some points equally spaced in $\log \ell$). The long-dashed line is the Gaussian which best fits the points (although not very well). The short-dashed line is the much better-fitting Lorentzian (see Eq. (1)).

The choice of fitting function was motivated by the need for simplicity and the requirement of a gradual rise into the Doppler peaks at $\ell \simeq 250$ (in CDM models the main Doppler peak occurs at $\ell \simeq 220$, with subsidiary peaks at higher $\ell$ – we have fitted *all* the Doppler peaks with a 'one-size fits all' function). Our chosen form will not be a good approximation for experiments that lie in the range of $\ell$ where the accurate Doppler peaks are dropping off. However, all such experiments are currently only giving upper limits, and none provide very tight constraints on CDM-like models (see Fig. 2). Our approach will also tend to slightly overestimate the power in experiments that have some sensitivity around the first Doppler trough

(causing our fit to prefer a *lower* peak height). However, most experimental points lie on the rise of the main Doppler peak, where our fit is extremely good.

To make contact with another possibility that has been discussed in the literature, we will also consider 'power law spectra', or Sachs-Wolfe fluctuations arising from non-flat power spectra. For these models[28,24,1],

$$C_\ell = C_2 \frac{\Gamma\left(\ell + \frac{n-1}{2}\right) \Gamma\left(2 + \frac{5-n}{2}\right)}{\Gamma\left(2 + \frac{n-1}{2}\right) \Gamma\left(\ell + \frac{5-n}{2}\right)}. \qquad (2)$$

## 3. Different Experimental Results

In order to use the results from several experiments at once, we need to convert them into a consistent system. The most straightforward and robust datum from each experiment is the total measured power. A simple parameterization of this power, integrated across the window function (bandpass) of the experiment is given by the amplitude of a *flat* power spectrum* $\ell(\ell+1)C_\ell = \text{constant} = (24\pi/5) \, Q_{\text{flat}}^2$, required to reproduce the measured power:

$$\text{Power} = \frac{1}{4\pi} \sum_{\ell=2}^{\infty} \frac{24\pi}{5} \frac{(2\ell+1)}{\ell(\ell+1)} Q_{\text{flat}}^2 W_\ell. \qquad (3)$$

where $W_\ell$ is the (diagonal) window function of the experiment (see e.g. ref. 30). The constants in this expression have been chosen so that $Q_{\text{flat}}$ has the same meaning as the more familiar[31] $Q_{\text{rms-PS}}$ for $n = 1$. Note also that $Q_{\text{flat}}$ is chosen to be independent of the observer's conventions for $\Delta T$.

Our estimated values for different experiments are listed in Table 1 and shown in Fig. 2. Each data point represents a fit for the amplitude of a flat spectrum convolved with the window function of each experiment. The vertical error bars are $1\sigma$ errors on this power, while the horizontal lines show the widths of the window functions at half peak height (and so should not be regarded as error bars), with some liberties taken with the way we have plotted COBE and FIRS. For the error bars on the 'power', we have taken them to be symmetric in $Q_{\text{flat}}$, or the same quantity as a '$\Delta T/T$' measurement. Many published experimental error bars are quoted as symmetric, others we have symmetrized. The most sensible way of plotting the experimental results is thus to choose a linear $y$-axis, which keeps the error bars symmetric and allows you to judge the significance of each detection. On the other hand it makes more sense to plot a theory in terms of the computed $(\Delta T/T)^2$ quantity. However, this squaring exaggerates the size of the experimental error bars. A logarithmic $y$-axis allows you to slide a theory up and down to find the best-fit normalization, but again makes it hard to interpret error bars. These considerations make it difficult to usefully plot data and theories together!

---

* We have tried to select the most transparent notation for this amplitude, bearing in mind that it is the expected quadrupole you would get if you put a *flat* spectrum through the experimental window function (and not a standard CDM, or even pure Sachs-Wolfe spectrum, which would depend on scale). The idea of quoting the power through the window has also been discussed by Bond[9,29].

**Table 1:** Summary of scales and predictions for current experiments. The parameters $\ell_0$, $\ell_1$ and $\ell_2$ are the peak and the lower and upper half-peak points of the window function, respectively. $Q_{\rm flat}$ is the best-fit amplitude for a flat spectrum through the window function, quoted at the quadrupole scale. The error bars are $\pm 1\sigma$.

| Experiment | | $\ell_0$ | $\ell_1$ | $\ell_2$ | $Q_{\rm flat}(\mu K)$ |
|---|---|---|---|---|---|
| COBE | [33] | – | – | 18 | $19.9 \pm 1.6$ |
| FIRS | [35] | – | – | 30 | $19 \pm 5$ |
| Ten. | [7] | 20 | 13 | 30 | $26 \pm 6$ |
| SP91 | [48] | 66 | 32 | 109 | $14 \pm 5$ |
| SK93 | [42] | 71 | 44 | 102 | $21 \pm 7$ |
| Pyth. | [37] | 73 | 50 | 107 | $37 \pm 12$ |
| ARGO | [44] | 107 | 53 | 180 | $25 \pm 6$ |
| IAB | [45] | 125 | 60 | 205 | $61 \pm 27$ |
| MAX–2 ($\gamma$ UMi) | [50] | 158 | 78 | 263 | $74 \pm 31$ |
| MAX–3 ($\gamma$ UMi) | [39] | 158 | 78 | 263 | $50 \pm 11$ |
| MAX–4 ($\gamma$ UMi) | [51] | 158 | 78 | 263 | $48 \pm 11$ |
| MAX–3 ($\mu$ Peg) | [38] | 158 | 78 | 263 | $19 \pm 8$ |
| MAX–4 ($\sigma$ Her) | [52] | 158 | 78 | 263 | $39 \pm 8$ |
| MAX–4 ($\iota$ Dra) | [52] | 158 | 78 | 263 | $39 \pm 11$ |
| MSAM2 | [41] | 143 | 69 | 234 | $40 \pm 14$ |
| MSAM3 | [41] | 249 | 152 | 362 | $39 \pm 12$ |

There are a number of issues that arise in dealing with these data. Space prevents us from going into every detail, but below we discuss some of the main points. We have chosen to use only quoted detections (see Table 1), and to neglect experiments that have given upper limits (we plot three smaller-scale upper limits in Fig. 2, but do not use them in our fit). Generally the error bars on these upper limits are large enough that they would not affect our results[4]. We have represented the COBE experiment by a single point, when it in fact has information for a range of $\ell$ [32,33]. This is also true of the FIRS experiment[34,35]. Furthermore, since we are fitting to the $Q$ for a flat model, it is appropriate for us to take the COBE[33] and FIRS[35] results for an $n = 1$ spectrum. For these experiments we plot the data point mid-way (logarithmically) to the half power $\ell$ of the window function. The position of the points on the plot is a visual aid to the scale these experiments probe; in analyzing the data we use the full window function as described below.

There is also the question of the 'sample variance'[36] of the experiments, i.e. the fact that looking at only part of the sky affects the error bar when comparing with a theoretical model. If the correlations between experimental data points are included in a proper statistical analysis (which is the case for all the numbers we consider) then the final result will have both the cosmic and sample variance fully included. We have also included the quoted calibration error for all the numbers where it was not included in the original papers, by adding it in quadrature to the quoted error bars.

In computing $Q_{\text{flat}}$ for the experiments, we have used whatever data were publicly available. For Python[37] and MAX–3[38,39], we fitted the given data to a flat spectrum. For COBE and FIRS, the results quoted by the groups come directly from a fit to a form like Eq. (3). For the Tenerife[40,7] experiment we used the published Harrison-Zel'dovich normalization.

For the MSAM[41], Saskatoon/SK93[42], ARGO[43,44] and the Italian Antarctic Base[45] data, we scaled from the quoted results for a Gaussian Autocorrelation Function (GACF). By this we mean that we calculated the power represented by a GACF of the quoted amplitude, and matched it to our chosen measure of power:

$$Q_{\text{flat}}^2 \sum_2^\infty \frac{(2\ell+1)}{5} \frac{6 W_\ell}{\ell(\ell+1)} = \frac{1}{2} \sum_2^\infty (2\ell+1) C_0 \theta_c^2 e^{-1/2 \ell(\ell+1) \theta_c^2} W_\ell. \quad (4)$$

Although the GACF is not a good approximation for the sky fluctuations, the power estimated in this way will not be too far wrong; it also has the virtue of taking into account the sample variance, since the correlations are done at least approximately correctly (see e.g. ref. 4 for a comparison of GACF and CDM correlation functions with and without beam chopping). Note also that by using $Q_{\text{flat}}$ as our measure of power, the normalization of the window functions cancels in the conversion $C_0^{1/2} \to Q_{\text{flat}}$ and in our fit. Further details of this conversion can be found in our companion paper[46].

We also chose to use the results from the full MSAM data, since we see no compelling reason to identify any point sources in the scans (see ref. 47). If these 'sources' are in fact removed, the two points (for the single- and double-differencing analyses) move roughly to the positions of the lower error bars. This would decrease the significance of our final results, although not greatly.

For SP91 we use the 13-pt scan[48] and neglect the upper limit from the 9-pt scan[49], which has little constraining power[4]. We performed a fit to the quoted data including correlations. It appears that combining the data from the two scans would lead to a higher value than for the 13-pt data alone[9], so choosing only the 13-point scan is conservative in this regard.

The MAX experiment also presents complications, since there are now six available scans. The $\mu$ Peg data set is known to be contaminated and gives a lower signal than the others, also there is a question as to whether the error bar accounts for all the error in the dust subtraction. However in the absence of a new analysis of these data, we decided that it would be a statistically dangerous thing to remove a data set just because it seemed disparate. We adopted a general philosophy that the systematic errors in these experiments are potentially large (so that some results will turn out to have been dominated by some non-primordial effect), but that once there are enough pieces of experimental data then a few discrepant points should carry little weight. For the earlier (flight 2)[50] and the newer (flight 4)[51,52] MAX data, we scaled from the quoted GACF results, after symmetrizing the rather positively skewed error bars. For the third flight 'GUM' (i.e. $\gamma$ UMi) scan we use the results of ref. 6. For the purposes of plotting and fitting the data we add the results in quadrature, except for the discrepant $\mu$ Peg scan, which we plot separately. This is statistically reasonable since the all the results except $\mu$ Peg are consistent[52]. We lose nothing by averaging the consistent data sets at the same angular scale; however including the $\mu$ Peg scan in the average would be bad, since it would make the final fit appear better than it should. In fact the window functions for flights 2 through 4 of the MAX experiment are not identical, but we feel that they are similar enough that we can treat them with an 'average' window function without introducing significant error. We have attempted to deal with the problem of the beam-size varying with frequency for the fourth flight, by carefully calculating the separate $C_0^{1/2}$ to $Q_{\text{flat}}$ conversions for each frequency and then taking a weighted average of the results. A full comparison of the MAX data with models still needs to be done, but we doubt that the final result will differ by more than a few $\mu$K from the values we have adopted.

We present our estimates for the power ($Q_{\text{flat}}$ in $\mu$K) obtained from each set of experimental data in Table 1. We also list the peak and half peak points of the window function for each experiment. These data are plotted in Fig. 2, along with three of the tightest ($2\sigma$) upper limits at somewhat smaller angular scales. The upper limits are for the White Dish[53], OVRO (NCP program[54]) and ATCA[55], and have been calculated from the quoted GACF results. In Fig. 2, if the power spectrum was *actually* Harrison-Zel'dovich, then the points would scatter about a horizontal line on this plot. The fact that there appears to be a trend for the degree-scale experiments to lie above such a line is what we will examine next.

### 4. How High is the Doppler Peak?

Taking the experimental measures of power from Table 1 and the toy-model power spectrum of Eq. (1), we can employ a likelihood analysis using the data to fit the two parameters, i.e. the overall normalization and the height of the Doppler peak. We should point out that we *do not* just fit a curve through the points of Fig. 2, which would not be an accurate procedure. The proper method is to convolve the fitting function with the window function for each experiment and compare the power with that obtained from the flat spectrum. This defines the set of *predicted* $Q$'s for each theory, which can be compared with the data in Table 1. We emphasize this so that the reader understands that the $Q_{\text{flat}}$ vs. $\ell$ plot should be regarded as a visual aid and is *not* directly what we used to calculate the fits.

A cursory glance at Fig. 2 is enough to realize that there are at least one or two of the data points which are not in very good agreement with the others, for *any* model, given the quoted error bars. It may be considered that this is evidence for non-Gaussian fluctuations (see e.g. ref. 56), but we believe it is more likely to be telling us that these are hard experiments, which have to contend with many technical difficulties, systematic errors (e.g. calibration errors at the 10% level) and possible foreground contamination. One way to view the data points is that their error bars have been underestimated to some extent. However, we find that the best fit Doppler model is allowed at the 90% CL. This shows that there is in fact no strong statistical reason to increase the error bars on the points – the data are more consistent than many people have been suggesting.

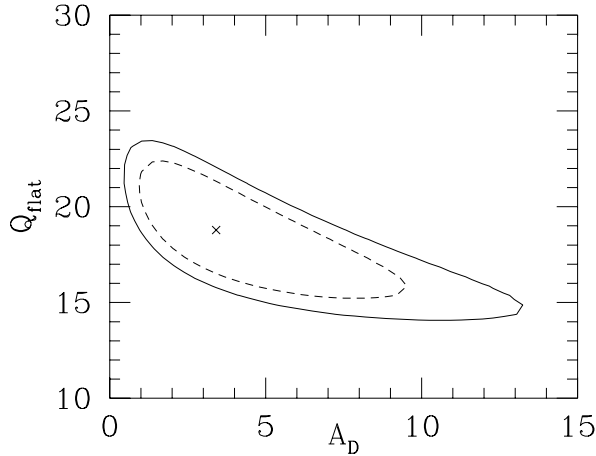

**Figure 3:** Contours of $\chi^2$ for a fit to the data of Table 1 using our phenomenological curve from Eq. (1). The cross marks the best fit ($Q_{\text{flat}} = 19\,\mu\text{K}$, $A_D = 3.4$), while the contours mark 68% and 95% CL regions for the fit parameters.

Having established that a 'good' fit exists, we will henceforth, following the Bayesian approach, ignore the goodness of fit and scale our likelihood function to unit integrated probability. A contour plot of the allowed range in $Q_{\text{flat}}$ and $A_D$ is shown in Fig. 3. The power spectrum normalization is well fixed by large scale measurements. To focus on the Doppler peaks, we show in Fig. 4 the 'marginal likelihood' or $\mathcal{L}(C_2^{1/2}, A_D)$ integrated over $C_2^{1/2}$ (with uniform prior). The best fit is $A_D \simeq 3$ and the mean $\simeq 4$. We can also ask what is the best fit for a straight power law and for a tilted CDM spectrum. The marginal likelihoods are shown in Fig. 5, both scaled to have unit area. The best fitting power law model is a worse fit than our best fitting "Doppler" model, and is (just) ruled out at the 95 %CL.

## 5. Conclusions

So what does this mean? The existence of the Doppler peaks first of all implies that reionization was relatively unimportant for CMB anisotropies, i.e. that the Thomson scattering optical depth since the Universe became ionized is not very significant (see later). This is perhaps not unexpected in cosmological models like CDM, which have little small-scale power to collapse and reheat the IGM at early epochs. But the existence of the Doppler peaks is also a confirmation of a quite fundamental theoretical prediction. The Doppler peak(s) occurs at approximately the position and size predicted many years ago for models with a few percent baryons with dark matter added to make up critical density, and a roughly scale-invariant primordial power spectrum. There *were* in fact baryons at $z \sim 1000$! They were once tightly coupled and oscillating with the photons, up until recombination at $z \simeq 1000$. And the amplitude of the oscillations is consistent with the nucleosynthesis constraint on $\Omega_B$ of a few percent.

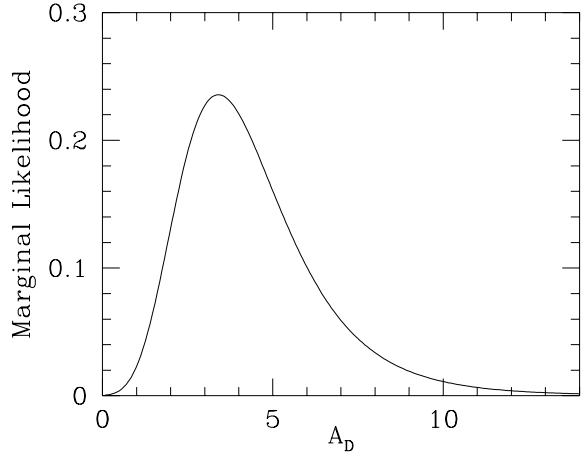

**Figure 4:** The marginal likelihood, or likelihood integrated over $C_2^{1/2}$, as a function of $A_D$ for our fitting form in Eq. (1). The likelihood has been normalized to have unit area. A fit with no Doppler peak ($A_D = 0$) is ruled out at 95% CL.

What about 'non-standard' scenarios? In texture models the peaks are generally absent[56], both because of the lower velocities in such models and because they generally invoke reionization. However, reionization is not necessary in such models. Without it we would probably expect Doppler peaks, although no explicit calculation has so far been done[57]. Certainly the microwave background would be expected to be highly non-Gaussian on such angular scales (roughly the horizon size at last scattering) in texture models. It seems that the similarity in fluctuations in the three dust-free regions scanned by the MAX experiment may already be evidence against such non-Gaussian models. The predictions of defect models will also depend on the choice of defect, for example with cosmic strings as seeds the anisotropies would not necessarily be non-Gaussian until much smaller angular scales[58]. But obviously time will tell how these models fare!

Also, the fact that our fitting formula, which has a plateau for low $\ell$ before rising into the peak, manages to pass through much of the data could be taken as evidence against an open universe isocurvature model (such as BDM/PIB/PBI). This model rises quickly into the Doppler peaks and is not flat at large scales[59,27]. The model is rather hard to rule out, since it has so many free parameters ($\simeq 8$), but detailed comparison with CMB results in the near future will be a critical test.

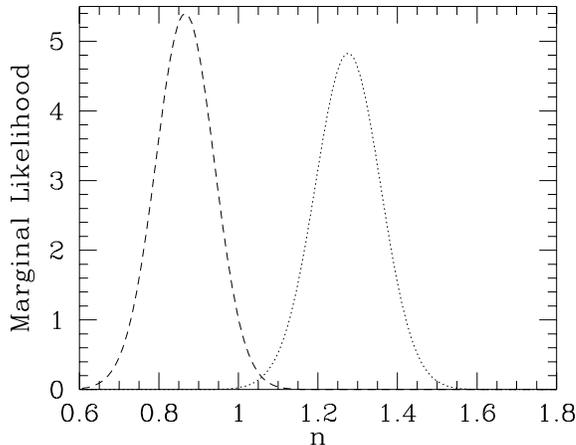

**Figure 5:** The marginal likelihood, or likelihood integrated over $C_2^{1/2}$, as a function of $n$ for an $n \neq 1$ Sachs-Wolfe spectrum (dotted) and for a tilted CDM model with $\Omega_B = 0.10$ (dashed). Both likelihoods have been normalized to unit area.

In the context of an inflationary dark matter-based theory, we can also ask for information on another parameter: the primordial spectral slope, $n$. Unambiguously determining this parameter is well beyond the scope of this work, although the 'peak' in the data at $\ell \simeq 250$, in combination with the COBE measurement, allows us to put a lower limit on $n$. Such a lower limit is most conservative if we ignore the possibility of gravity waves. From Big Bang Nucleosynthesis we 'know' that $\Omega_B$ cannot be arbitrarily large. In fact a value of $\Omega_B$ as large as 10% seems unlikely. Since the Doppler peak height increases with $\Omega_B$, a lower limit on the tilt of such a model is a *conservative* lower limit for any model with a more reasonable value of $\Omega_B$. From Fig. 4 we see that the CMB data alone appear to require $n > 0.7$ at the 95% CL, even for such a high $\Omega_B$, which is competitive with combinations of large scale CMB and LSS data ($n \gtrsim 0.7$, e.g. ref. 60).

As an alternative to limiting $n$, we can obtain a crude limit on the ionization history of the universe. Recall[61] that in a reionized universe the degree scale temperature anisotropies are reduced, relative to those in the standard model, by a factor $e^{-\tau}$. If we compare this with the amount that degree scale anisotropies are reduced in a tilted model, $\ell^{(1-n)/2}$, we can use our limit above to find $\tau \lesssim 0.7$. Assuming, as above, that $\Omega_B = 0.10$ and with full ionization ($x_e = 1$) at the present epoch, we find that the universe had to have been neutral between redshifts $\simeq 50$ and 1000.

Knowledge of other cosmological parameters will affect our fits to some extent. For example if $\Omega_0 < 1$, $\Lambda > 0$, $n \neq 1$, $T/S > 0$ etc., then the height of the Doppler peaks will change relative to a COBE normalization. However, the indications are that none of these effects are so important as to invalidate our results. More rigorous and constraining fits could clearly be done, but we feel that the current data do not warrant a complicated multi-parameter fit. In particular we have avoided the temptation to derive any specific cosmological parameter instead of our phenomenological amplitude $A_D$. However, we cannot entirely resist saying that $A_D = 3$–$4$ would correspond to $\Omega_B = 1$–$3\%$ for a standard CDM model with $h = 1/2$. But this result is really quite meaningless as a measurement of $\Omega_B$ since it depends sensitively on what is assumed for the other cosmological parameters.

The potential for doing cosmology with the spectrum of microwave background anisotropies is finally being realized. The COBE detection allowed us to normalize models, and now the detections at degree-scales are indicating the reality of Doppler peaks of some sort. The task for the immediate future will be to determine exactly where and how high they are. The position of the main Doppler peak gives an exciting possibility of being able to 'prove' that the Universe is open; if it is at $\ell \simeq 500$ we will have to take seriously the idea that $\Omega \simeq 0.2$ say (although it will be more difficult to 'prove' that $\Omega = 1$ if it is at $\ell \simeq 220$). The height of the Doppler peaks will be an important constraint on a combination of cosmological parameters, perhaps mainly $\Omega_B$, although determining the size of the subsidiary peaks may give us information on the Hubble parameter. In the more distant future, after we have accurately charted the Doppler peaks and have other cosmological evidence for constraining $H_0$, $\Lambda$ and the reionization history, there is a chance of being able to detect a component of gravity waves as well as measuring $n$ for the scalars. This will lead to some fundamental constraints on the physics of the early Universe. However, for now the determination of more 'classical' cosmological parameters is an ambitious enough goal!

## Acknowledgements

We would like to thank Naoshi Sugiyama for providing us with the $C_\ell$'s for CDM, and also Wayne Hu and him for many useful conversations. Dick Bond provided some helpful comments and Joe Silk provided constant enthusiasm. This work was supported in part by grants from the NSF. M.W. acknowledges the support of an SSC fellowship from the TNRLC.

**Figure 2 (colour plate):** The 'power' in each experiment as a function of scale (multipole $\ell \sim \theta^{-1}$). Detections from 10 experiments are plotted, with the MSAM and MAX experiments each represented by two separate points. The vertical error bars are $\pm 1\sigma$, while the horizontal bar represents the half power range of the experimental window function. There are also three smaller-scale upper limits plotted at the $2\sigma$ level. It is possible to sense a general 'upness' in the area around $\ell \simeq 100$, which we claim is evidence for a Doppler peak in the radiation power spectrum.

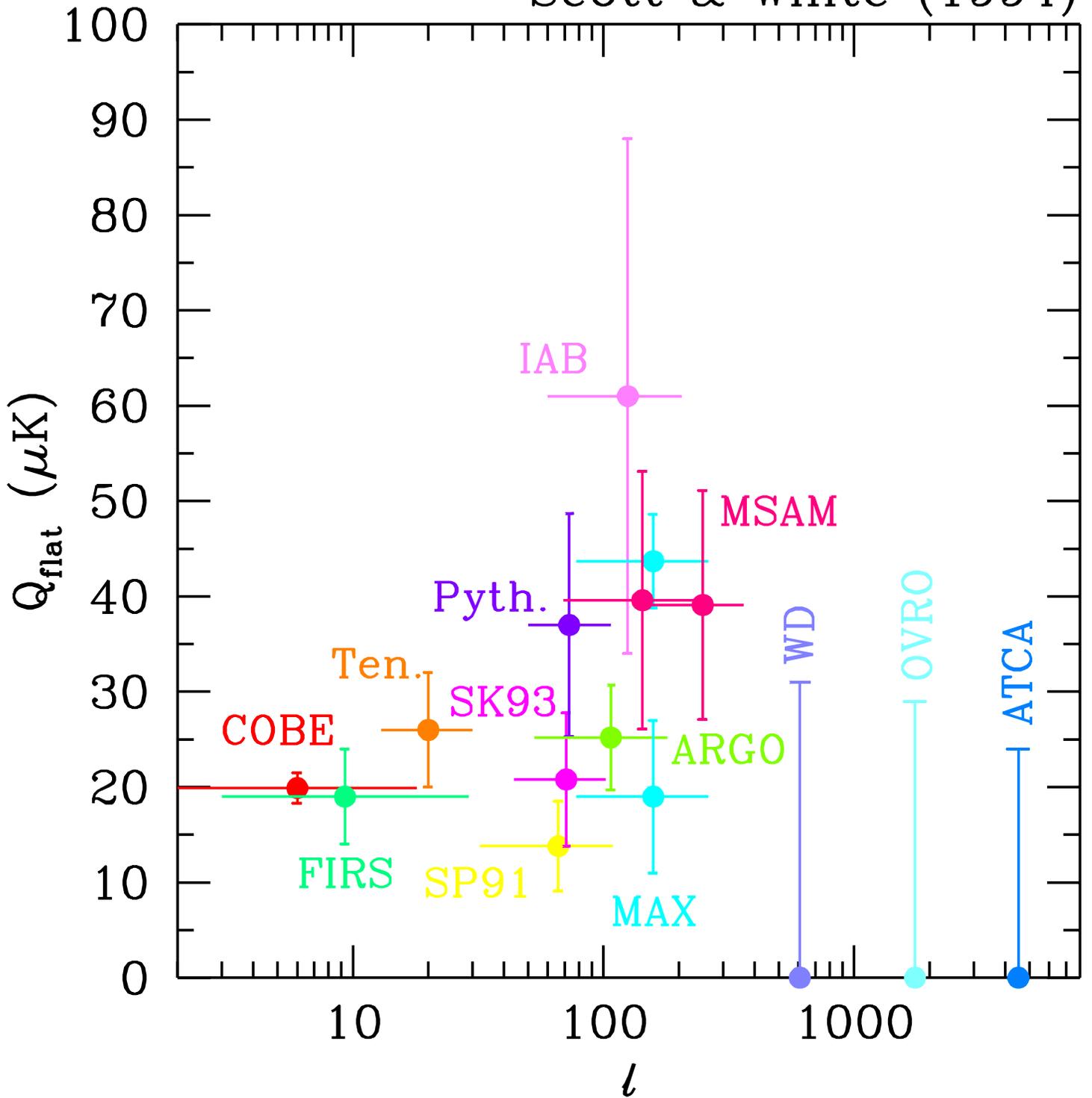